\documentclass[english,11pt]{article}
\usepackage[T1]{fontenc}
\usepackage[utf8]{inputenc}
\usepackage{array}
\usepackage{multirow}
\usepackage{amsmath}
\usepackage{amssymb}
\usepackage{graphicx}

\makeatletter

\providecommand{\tabularnewline}{\\}

\usepackage{amssymb}
\usepackage{latexsym}
\usepackage{epsfig}
\usepackage{float}
\usepackage{dsfont}

\newcommand{\be}{\begin{equation}}
\newcommand{\ee}{\end{equation}}

\usepackage[numbers,sort&compress]{natbib}
\usepackage{hyperref}
\RequirePackage{doi}

\@ifundefined{showcaptionsetup}{}{%
 \PassOptionsToPackage{caption=false}{subfig}}
\usepackage{subfig}
\makeatother

\usepackage{babel}
\begin{document}
{}~ \hfill\vbox{\hbox{CTP-SCU/2019012}}\break
\vskip 3.0cm
\centerline{\Large \bf  Strong Cosmic Censorship for a Scalar Field in
 }
\vspace*{1.5ex}
\centerline{\Large \bf  
a Born-Infeld-de Sitter Black Hole }

\vspace*{10.0ex}
\centerline{\large Qingyu Gan, Guangzhou Guo, Peng Wang and Houwen Wu}
\vspace*{7.0ex}
\vspace*{4.0ex}
\centerline{\large \it Center for Theoretical Physics, College of Physics}
\centerline{\large \it Sichuan University}
\centerline{\large \it Chengdu, 610064, PR China} \vspace*{1.0ex}
\vspace*{4.0ex}

\centerline{qygan@stu.scu.edu.cn, guangzhouguo@stu.scu.edu.cn, pengw@scu.edu.cn, iverwu@scu.edu.cn}
\vspace*{10.0ex}
\centerline{\bf Abstract} \bigskip \smallskip
It has been shown that the Strong Cosmic Censorship (SCC) can be violated by a scalar field in a near-extremal Reissner-Nordstrom-de Sitter black hole. In this paper, we investigate the Strong Cosmic Censorship in a Born-Infeld-de Sitter black hole by a scalar perturbation field with/without a charge. When the Born-Infeld parameter $b$ becomes small, the nonlinear electrodynamics effect starts to play an important role and tends to rescue SCC. Specifically, we find that the SCC violation region decreases in size with decreasing $b$. Moreover, for a sufficiently small $b$, SCC can always be restored in a near-extremal Born-Infeld-de Sitter black hole with a fixed charge ratio.

\vfill 
\eject
\baselineskip=16pt
\vspace*{0.0ex}
\tableofcontents

\section{Introduction}

A black hole formed from gravitational collapse could possess a curvature
singularity. If a time-like singularity is formed, the undetermined
initial data on it would cause the breakdown of determinism of general
relativity. On the other hand, it is well known that there exist some
solutions of the Einstein field equations admitting the presence of
time-like singularities, e.g., Kerr-Newman black holes, Reissner-Nordstrom
black holes. To rescue the predictability of general relativity, Penrose
proposed a conjecture, namely the Strong Cosmic Censorship (SCC),
asserting that the maximal Cauchy development of physically acceptable
initial conditions is locally inextendible as a regular manifold \cite{Penrose:1969pc,Hawking:1969sw,Penrose:1964wq}.
Consequently, when the initial data is perturbed outside of a black
hole, whether SCC holds true crucially depends on the extendibility
of the perturbation (e.g., the metric and other fields) at the Cauchy
horizon, which encloses the time-like singularity.

Nevertheless, there are some subtleties of describing the extension
of the perturbation at the Cauchy horizon, and hence several formulation
versions of SCC have been proposed. For example, the perturbation
can not be $C^{r}(r\in N)$ smooth at the Cauchy horizon in the $C^{r}$
version of SCC \cite{Luk:2017jxq,Luk:2017ofx}. However, since weak
solutions can have many important physical applications in which $C^{r}$
smooth solutions are not available, a more appropriate way to characterize
the extendibility is whether the perturbation being inextendible as
a weak solution of the equations of motion. This observation then
leads to the Christodoulou version of SCC \cite{Christodoulou:2008nj}.
In other words, if SCC is violated in the Christodoulou version, the
perturbation belongs to the Sobolev space $H_{loc}^{1}$, where the
first derivative of the perturbation is locally square integrable
and, roughly speaking, has finite energy at the Cauchy horizon.

The possible singular behavior of a perturbation at the Cauchy horizon
comes from the mass-inflation mechanism, associated with the exponential
amplification due to the blue shift effect \cite{Chambers:1997ef,Dafermos:2003wr,Poisson:1990eh,Ori:1991zz,Hod:1998gy,Brady:1995ni}.
However, there is another mechanism competing with the mass-inflation
mechanism to invalidate SCC: the time-dependent remnant perturbation
decaying outside of the black hole. When the perturbation decays slowly
enough, SCC could be valid. In fact, a perturbation in an asymptotically
flat black hole satisfies an inverse power law decay \cite{Price:1971fb,Dafermos:2014cua,Angelopoulos:2016wcv},
which ensures the mass-inflation mechanism is strong enough to render
the Cauchy horizon unstable upon perturbation \cite{Dafermos:2003wr,Dafermos:2012np}.
On the other hand, it was observed that a remnant perturbation can
exponentially decay in a black hole in asymptotically dS space-time
\cite{2007arXiv0706.0350B,Dyatlov:2013hba,Dyatlov:2011jd,Hintz:2016gwb,Hintz:2016jak,Berti:2009kk,Konoplya:2011qq,Brady:1996za},
which implies that the perturbation might have chance to decay fast
enough to violate SCC. More precisely, it showed that, for an asymptotically
dS black hole, the competition between the the mass inflation and
remnant decaying can be characterized by \cite{Costa:2014aia,Costa:2014yha,Costa:2014zha,Hintz:2015jkj,Kehle:2018upl,Cardoso:2017soq}
\begin{equation}
\beta\equiv\frac{\alpha}{\kappa_{-}},\label{eq:beta}
\end{equation}
where $\kappa_{-}$ denotes the surface gravity at the Cauchy horizon,
and $\alpha$ is the spectral gap representing the distance from the
real axis to the lowest-lying Quasi-Normal Mode (QNM) on the lower
half complex plane of frequency. Note that $\beta>1/2$ can lead to
a potential violation of the Christodoulou version of SCC.

Recently, the validity of the Christodoulou version of SCC has been
explored in asymptotically dS black holes by computing $\beta$ for
various perturbation fields \cite{Cardoso:2017soq,Mo:2018nnu,Dias:2018ufh,Cardoso:2018nvb,Hod:2018dpx,Ge:2018vjq,Destounis:2018qnb,Rahman:2018oso,Liu:2019lon,Dias:2018etb,Luna:2018jfk,Dias:2018ynt,Gwak:2018rba,Guo:2019tjy}.
In particular, a massless neutral scalar perturbation field in a Reissner-Nordstrom-de
Sitter (RN-dS) black hole was considered in \cite{Cardoso:2017soq},
and it was proven that SCC is violated in the near-extremal regime.
Since the charge matter is necessary for the formation of a charged
black hole by gravitational collapse, the analysis was then extended
to a charged scalar field in a RN-dS black hole \cite{Cardoso:2018nvb,Mo:2018nnu,Dias:2018ufh},
which showed that, in the highly extremal limit, there always exists
a region in parameter space where SCC is violated. Although it was
claimed in \cite{Hod:2018dpx} that SCC would be saved for sufficiently
large scalar field mass and charge, the existence of arbitrarily small
oscillations of $\beta$ around $\beta=1/2$ was observed in a sufficiently
near-extremal black hole \cite{Dias:2018ufh}. These oscillations
were dubbed as ``wiggles'', which result from non-perturbative effects
and can lead to a violation of SCC for an arbitrary large scalar field
charge . Later, SCC in a RN-dS black hole was also discussed in the
context of the Dirac perturbation field \cite{Ge:2018vjq,Destounis:2018qnb}
and higher space-time dimensions \cite{Rahman:2018oso,Liu:2019lon},
where there still exists some room for the violation of SCC. Considering
smooth initial data, the violation of SCC becomes more severe for
the coupled linearized electromagnetic and gravitational perturbations
in a RN-dS black hole \cite{Dias:2018etb}. In \cite{Luna:2018jfk},
the authors proved that nonlinear effects could not save SCC from
being violated for a near-extremal RN-dS black hole. On the other
hand, SCC is always respected for the massless scalar field and linearized
gravitational perturbations in a Kerr-dS black hole \cite{Rahman:2018oso,Dias:2018ynt}.

Taking quantum contributions into account, nonlinear corrections are
usually added to the Maxwell Lagrangian, which gives the nonlinear
electrodynamics (NLED). Among various NLED, Born-Infeld (BI) electrodynamics,
which was first introduced to smooth divergences of the electrostatic
self-energy of point charges, has attracted considerable attention
in the literature. Furthermore, BI electrodynamics can come from the
low energy limit of string theory and encodes the low-energy dynamics
of D-branes \cite{Gibbons:2001gy}. The BI black hole solution in
(A)dS space was first obtained in \cite{Dey:2004yt,Cai:2004eh}. Since
then, various aspects of BI black holes have been extensively investigated,
e.g., the thermodynamics and phase structure \cite{Fernando:2006gh,Gunasekaran:2012dq,Zou:2013owa,Zeng:2016sei,Wang:2018xdz,Wang:2019kxp},
the holographic models \cite{Jing:2010zp,Baggioli:2016oju,Kiritsis:2016cpm,Tao:2017fsy,Guo:2017bru,Cremonini:2017qwq,Wang:2018hwg,Gan:2018utc}.
Specifically, the Weak Cosmic Censorship (WCC) has recently been studied
in a BI black hole \cite{Wang:2019dzl,Hu:2019rpw}, where it was found
that there may exist some counterexamples to WCC.

Until now, the charge sector of testing the Christodoulou version
of SCC has been confined to Maxwell's theory of electrodynamics. Little
is known about the NLED effect on the validity of SCC. In this paper,
we investigate the Christodoulou version of SCC for a scalar perturbation
field propagating in a BI-dS black hole. Our results show that the
NLED effect tends to alleviate the violation of SCC in the near-extremal
regime. Especially, for a near-extremal BI-dS black hole with a fixed
charge ratio, SCC can always be saved as long as the NLED effect is
strong enough. Furthermore, the parameter region where SCC is violated
decreases in size when the NLED effect becomes stronger.

The rest of the paper is organized as follows. In Section \ref{sec:BI-dS-black-hole},
we briefly review the BI-dS black hole solution and obtain the parameter
region where the Cauchy horizon exists. In Section \ref{sec:3 Quasinormal-Mode},
we show how to compute the QNMs for a charged and massive scalar field
in a BI-dS black hole. In Section \ref{sec:Numerical-results}, we
present and discuss the numerical results in various parameter regions.
We summarize our results in the last section. For simplicity, we set
$16\pi G=c=1$ in this paper.

\section{BI-dS Black Hole}

\label{sec:BI-dS-black-hole}

In this section, we first review the BI-dS black hole solution and
then give the ``allowed'' region in the parameter space, in which
the Cauchy horizon exists. Consider a $\left(3+1\right)$-dimensional
Einstein-Born-Infeld action in the presence of the positive cosmological
constant $\Lambda$, which is given by 
\begin{equation}
S=\int d^{4}x\sqrt{-g}\left[\mathcal{R}-2\Lambda+4b^{2}\left(1-\sqrt{1+\frac{F^{\mu\nu}F_{\mu\nu}}{2b^{2}}}\right)\right],\label{eq:action}
\end{equation}
where $\mathcal{R}$ is the Ricci scalar curvature, $F_{\mu\nu}=\partial_{\mu}A_{\nu}-\partial_{\nu}A_{\mu}$
is the electromagnetic tensor field of a BI electromagnetic field
$A_{\mu}$, and the Born-Infeld parameter $b$ is related to the string
tension $\alpha'$ as $b=1/(2\pi\alpha^{\prime})$ \cite{Gibbons:2001gy}.
It is noteworthy that BI electrodynamics would reduce to Maxwell electrodynamics
in the limit of $b\rightarrow\infty$. So the NLED effect in BI electrodynamics
will become stronger for a smaller value of $b$. For the action (\ref{eq:action}),
a static spherically symmetric black hole solution was obtained in
\cite{Cai:2004eh,Dey:2004yt}:

\begin{equation}
ds^{2}=-f\left(r\right)dt^{2}+\frac{dr^{2}}{f\left(r\right)}+r^{2}(d\theta^{2}+\sin^{2}\theta d\varphi^{2}),\quad\mathbf{A}=A_{t}dt=-\frac{Q}{r}{}_{2}\mathcal{F}_{1}\left[\frac{1}{4},\frac{1}{2},\frac{5}{4},-\frac{Q^{2}}{b^{2}r^{4}}\right]dt,\label{eq:metric and gauge potential}
\end{equation}
with the blackening factor

\begin{equation}
f\left(r\right)=1-\frac{\Lambda r^{2}}{3}-\frac{M}{8\pi r}+\frac{2b^{2}r^{2}}{3}\left(1-\sqrt{1+\frac{Q^{2}}{b^{2}r^{4}}}\right)+\frac{4Q^{2}}{3r^{2}}{}_{2}\mathcal{F}_{1}\left[\frac{1}{4},\frac{1}{2},\frac{5}{4},-\frac{Q^{2}}{b^{2}r^{4}}\right].\label{eq:fr}
\end{equation}
Here, $_{2}\mathcal{F}_{1}$ is the hypergeometric function, and $M$
and $Q$ are the mass and electric charge of the BI-dS black hole,
respectively. In the limit of $b\rightarrow\infty$, eqns. (\ref{eq:metric and gauge potential})
and (\ref{eq:fr}) recover the RN-dS black hole solution as expected.

\begin{figure}
\begin{centering}
\includegraphics[scale=0.8]{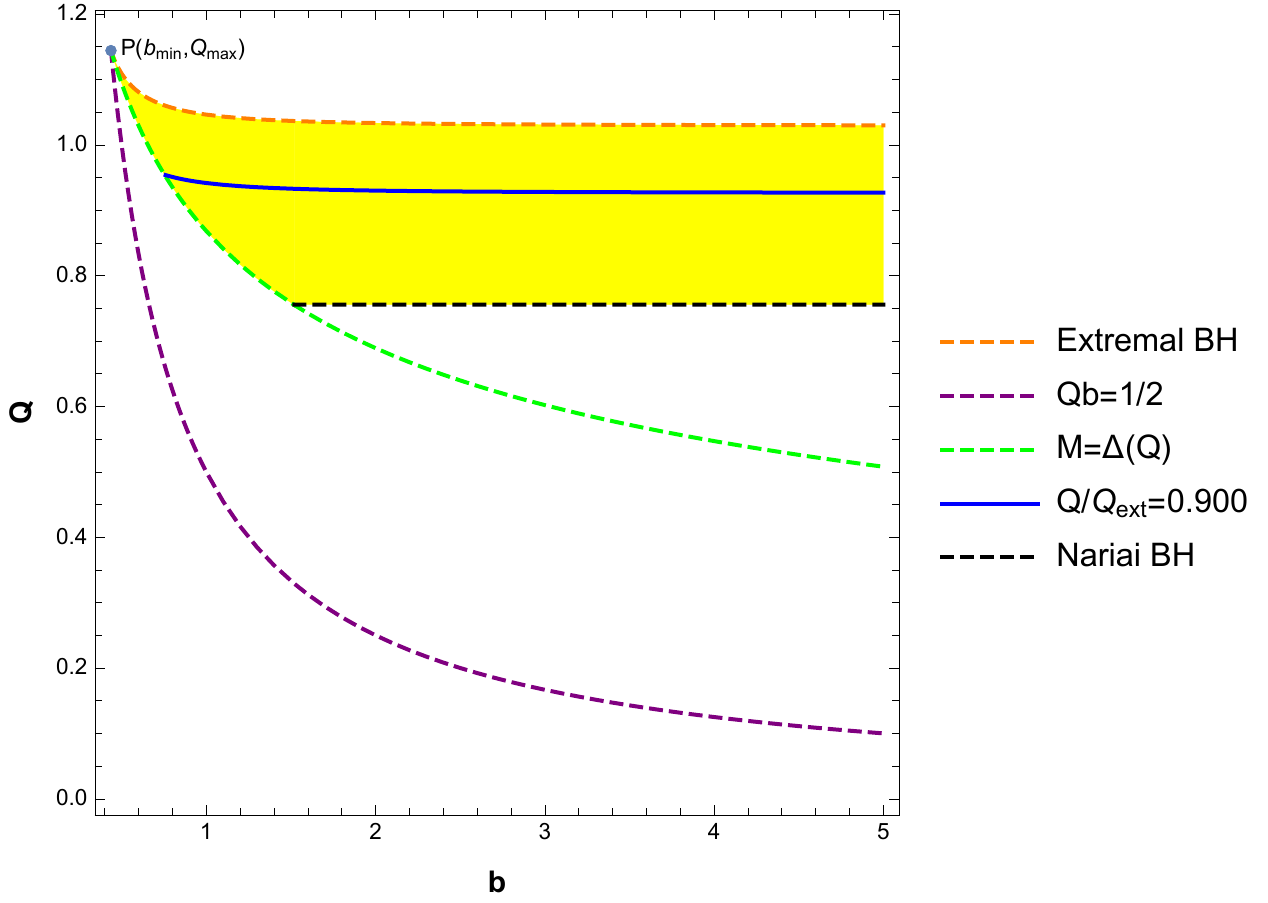}
\par\end{centering}
\caption{$M=16\pi$ and $\Lambda=0.14$. The BI-dS black hole in the yellow
region possesses three horizons. The dashed orange and black lines
represent the extremal black hole with $r_{-}=r_{+}$ and the Nariai
black hole with $r_{+}=r_{c}$, respectively. Note that the orange,
green and purple lines meet at the point $P$, which marks the minimum
value of $b$ and the maximum value of $Q$ in the yellow region.
The solid blue line corresponds to a near-extremal black hole with
the charge $Q=0.900Q_{\textrm{ext}}$, where $Q_{\textrm{ext}}$ is
the charge of the extremal black hole.}

\label{Region}
\end{figure}

A BI-dS black hole can possess one, two or three horizons depending
on the parameters $M$, $Q$, $\Lambda$ and $b$. The topology of
BI-dS black holes has been discussed in \cite{Fernando:2003tz,Fernando:2013uza}.
In this paper, we investigate SCC and hence focus on the BI-dS black
holes possessing three horizons, namely the Cauchy horizon $r_{-}$,
the event horizon $r_{+}$ and the cosmological horizon $r_{c}$.
To determine the number of roots of $f\left(r\right)$, we instead
consider $rf\left(r\right)$, which has the same positive roots as
$f\left(r\right)$, and find that

\begin{equation}
\frac{d\left(rf\left(r\right)\right)}{dr}=1+\left(2b^{2}-\Lambda\right)r^{2}-2b\sqrt{Q^{2}+b^{2}r^{4}}.\label{eq:derivative of rfr}
\end{equation}
In order for $rf\left(r\right)$ to have three positive roots, the
parameters must satisfy the following conditions:
\begin{itemize}
\item $0<\Lambda<2b^{2}$ and $bQ>\frac{1}{2}$: Since the BI-dS black hole
solution is asymptotically dS, one has $rf\left(r\right)\rightarrow-\infty$
in $r\rightarrow\infty.$ Thus, $rf\left(r\right)$ must have a local
minimum at $r=r_{\textrm{min}}$ and a local maximum at $r=r_{\textrm{max}}$,
which has $0<r_{\textrm{min}}<r_{\textrm{max}}$. The existence of
two extrema for $rf\left(r\right)$ requires that $d\left(rf\left(r\right)\right)/dr=0$
has two positive roots, which gives $0<\Lambda<2b^{2}$ and $bQ>\frac{1}{2}$.
\item $M<\varDelta(Q)$: At $r=0$, $rf\left(r\right)$ must be greater
than zero. As $r\rightarrow0\text{, one has }$
\begin{equation}
rf\left(r\right)=\frac{\varDelta(Q)-M}{8\pi}-\left(\frac{10}{3}bQ-1\right)r+\mathcal{O}(r^{2})\label{eq:limit-fr}
\end{equation}
with $\varDelta(Q)=\frac{8\pi}{3}\sqrt{\frac{b}{\pi}}Q^{3/2}\Gamma\left(\frac{1}{4}\right)^{2}$,
which leads to $M<\varDelta(Q)$.
\item $f\left(r_{\textrm{min}}\right)<0$: The local minimum value of $rf\left(r\right)$
at $r=r_{\textrm{min}}$ must be negative. When $f\left(r_{\textrm{min}}\right)=0$,
we have the extremal black hole solution with $r_{-}=r_{+}$. For
later use, $Q_{\textrm{ext}}$ denotes the charge of the extremal
black hole.
\item $f\left(r_{\textrm{max}}\right)>0$: The local maximum value of $rf\left(r\right)$
at $r=r_{\textrm{max}}$ must be positive. When $f\left(r_{\textrm{max}}\right)=0$,
we have the Nariai black hole solution with $r_{+}=r_{c}$, which
could only be calculated numerically \cite{Fernando:2013uza}.
\end{itemize}
The above conditions together give the allowed region in the parameter
space, in which a BI-dS black hole has three horizons. We also find
that there exist a lower bound $b_{\textrm{min}}$ on $b$ and an
upper bound $Q_{\textrm{max}}$ on $Q$ in the allowed region. In
fact, it can show that the boundaries $bQ=\frac{1}{2}$, $M=\varDelta(Q)$
and $f\left(r_{\textrm{min}}\right)=0$ can intersect at one point
in the $b$-$Q$ plane, which gives
\begin{equation}
b_{\textrm{min}}=\frac{2\sqrt{2\pi}\Gamma\left(\frac{1}{4}\right)^{2}}{3M},\quad Q_{\textrm{max}}=\frac{3M}{4\sqrt{2\pi}\Gamma\left(\frac{1}{4}\right)^{2}}.\label{eq:bmin qmax}
\end{equation}
We plot the allowed region and their boundaries in the $b$-$Q$ parameter
space in Fig. \ref{Region}, where $M=16\pi$ and $\Lambda=0.14$.

\section{Quasi-Normal Mode}

\label{sec:3 Quasinormal-Mode}

In this section, we discuss the QNMs for a charged and massive scalar
field in a BI-dS black hole. We first consider a scalar perturbation
of mass $\mu$ and charge $q$ governed by the Klein-Gordon equation
\begin{equation}
\left(\mathbf{D}^{2}-\mu^{2}\right)\Phi=0,\label{eq:KG eq.}
\end{equation}
where $\mathbf{D}$ denotes the covariant derivative $\mathbf{D}=\mathbf{\mathbf{\mathbf{\nabla}}}-iq\mathbf{A}$.
To facilitate our numerical calculation, we will use Eddington-Finkelstein
ingoing coordinates $\left(v,r,\theta,\phi\right)$ with $v=t+r_{*}$,
where $r_{*}$ is the tortoise coordinate defined as $dr_{*}=dr/f(r)$.
In addition, we choose an appropriate gauge transformation such that
$\mathbf{A}=A_{v}dv=-\frac{Q}{r}{}_{2}\mathcal{F}_{1}\left[\frac{1}{4},\frac{1}{2},\frac{5}{4},-\frac{Q^{2}}{\beta^{2}r^{4}}\right]dv$.
Since the BI-dS black hole solution is static and spherically symmetry,
a mode solution of eqn. (\ref{eq:KG eq.}) can have the separable
form

\begin{equation}
\Phi=e^{-i\omega v}Y_{lm}\left(\theta,\phi\right)\psi_{\omega l}\left(r\right),\label{eq:=00005CPhi}
\end{equation}
where $Y_{lm}\left(\theta,\phi\right)$ is the harmonic function of
the unit $2$-sphere. Plugging eqn. (\ref{eq:=00005CPhi}) into eqn.
(\ref{eq:KG eq.}), we obtain the radial equation
\begin{eqnarray}
0 & = & \left[r^{2}f\partial_{r}^{2}+\left(r^{2}f^{\prime}+2rf-2iqA_{v}r^{2}-2i\omega r^{2}\right)\partial_{r}\right.\nonumber \\
 &  & \left.-2i\omega r-2iqrA_{v}-iqr^{2}\partial_{r}A_{v}-l\left(l+1\right)-\mu^{2}r^{2}\right]\psi_{\omega l}\left(r\right)\label{eq:radial-equation}
\end{eqnarray}

where $f^{\prime}$ denotes $df(r)/dr$. One can perform the Frobenius
method to obtain the solutions near the event and cosmological horizons,
respectively. In fact, we define a new coordinate $x\equiv(r-r_{+})/(r_{c}-r_{+})$.
Near the event horizon, i.e., $x\rightarrow0$, $\psi_{\omega l}\left(r\right)$
has the ingoing and outgoing boundary solutions:
\begin{equation}
\psi_{\omega l}^{\textrm{ingoing}}\sim\textrm{const},\quad\psi_{\omega l}^{\textrm{outgoing}}\sim x^{i\frac{\omega+q\left.A_{v}\right|_{r=r_{+}}}{\kappa_{+}}}.
\end{equation}
And near the cosmological horizon, i.e., $x\rightarrow1$, $\psi_{\omega l}\left(r\right)$
also has the ingoing and outgoing boundary solutions:
\begin{equation}
\psi_{\omega l}^{\textrm{ingoing}}\sim\textrm{const},\quad\psi_{\omega l}^{\textrm{outgoing}}\sim(1-x)^{-i\frac{\omega+q\left.A_{v}\right|_{r=r_{c}}}{\kappa_{c}}}.
\end{equation}
Here $\kappa_{h}\equiv\left|f^{\prime}\left(r_{h}\right)\right|/2$
with $h\in\{+,-,c\}$ is the surface gravity at each horizon.

Imposing the ingoing boundary condition at the event horizon and the
outgoing boundary condition at the cosmological horizon on eqn. (\ref{eq:radial-equation})
selects a set of discrete frequencies $\omega_{ln}(n=1,2,\cdots)$,
called QNMs \cite{Berti:2009kk}. There are many analytic and numerical
ways to extract QNMs \cite{Berti:2009kk,Konoplya:2011qq}. Here we
we employ the Chebyshev collocation scheme and the associated Mathematica
package developed in \cite{Jansen:2017oag,Jansen,url-centra.tecnico.}.
We redefine field $\psi_{\omega l}$ adapted to our numerical scheme:
\begin{equation}
\psi_{\omega l}=\frac{1}{x}\left(1-x\right)^{-i\frac{\omega+q\left.A_{v}\right|_{r=r_{c}}}{\kappa_{c}}}\phi_{\omega l},\label{eq:redefined-phi}
\end{equation}
where the new field $\phi_{\omega l}$ becomes regular at both the
event and cosmological horizons. After the radial equation for $\phi_{\omega l}$
is obtained, we can use the Mathematica package to find a series of
QNMs, $\omega_{ln}$. The spectral gap $\alpha$ in eqn. (\ref{eq:beta})
is then given by $\alpha=\textrm{inf}{}_{ln}\left\{ -\textrm{Im(}\omega_{ln})\right\} $.

\section{Numerical Results}

\label{sec:Numerical-results}

\begin{table}
\begin{centering}
\begin{tabular}{|c|c|c|c|c|c|c|}
\hline 
$\Lambda$ & $b$ & $Q/Q_{\textrm{ext}}$ & $q$ & $l=0$ & $l=1$ & $l=10$\tabularnewline
\hline 
\hline 
\multirow{8}{*}{$0.02$} & \multirow{4}{*}{$0.5$} & \multirow{2}{*}{$0.991$} & $0$ & $0$ & $\pm0.586851-0.102835i$ & $\pm4.107680-0.099091i$\tabularnewline
\cline{4-7} \cline{5-7} \cline{6-7} \cline{7-7} 
 &  &  & $0.1$ & $0.015305+0.001095i$ & $0.686126-0.101651i$ & $4.204171-0.098960i$\tabularnewline
\cline{3-7} \cline{4-7} \cline{5-7} \cline{6-7} \cline{7-7} 
 &  & \multirow{2}{*}{$0.996$} & $0$ & $0$ & $\pm2.514067-0.411424i$ & $\pm17.649019-0.393928i$\tabularnewline
\cline{4-7} \cline{5-7} \cline{6-7} \cline{7-7} 
 &  &  & $0.1$ & $0.065124+0.004991i$ & $2.948696-0.400322i$ & $18.072264-0.392465i$\tabularnewline
\cline{2-7} \cline{3-7} \cline{4-7} \cline{5-7} \cline{6-7} \cline{7-7} 
 & \multirow{4}{*}{$10000$} & \multirow{2}{*}{$0.991$} & $0$ & $0$ & $-0.475688i$ & $\pm14.365381-0.491756i$\tabularnewline
\cline{4-7} \cline{5-7} \cline{6-7} \cline{7-7} 
 &  &  & $0.1$ & $0.057773+0.002229i$ & $0.032203-0.475118i$ & $-14.080016-0.491441i$\tabularnewline
\cline{3-7} \cline{4-7} \cline{5-7} \cline{6-7} \cline{7-7} 
 &  & \multirow{2}{*}{$0.996$} & $0$ & $0$ & $-0.789379i$ & $\pm23.969407-0.808962i$\tabularnewline
\cline{4-7} \cline{5-7} \cline{6-7} \cline{7-7} 
 &  &  & $0.1$ & $0.096356+0.003870i$ & $0.053708-0.788423i$ & $-23.488922-0.808825i$\tabularnewline
\hline 
\multirow{8}{*}{$0.06$} & \multirow{4}{*}{$0.5$} & \multirow{2}{*}{$0.991$} & $0$ & $0$ & $\pm0.686276-0.121754i$ & $\pm4.863532-0.116340i$\tabularnewline
\cline{4-7} \cline{5-7} \cline{6-7} \cline{7-7} 
 &  &  & $0.1$ & $0.039490+0.002376i$ & $0.817179-0.119918i$ & $4.990056-0.116174i$\tabularnewline
\cline{3-7} \cline{4-7} \cline{5-7} \cline{6-7} \cline{7-7} 
 &  & \multirow{2}{*}{$0.996$} & $0$ & $0$ & $\pm2.520908-0.414697i$ & $\pm17.902524-0.394830i$\tabularnewline
\cline{4-7} \cline{5-7} \cline{6-7} \cline{7-7} 
 &  &  & $0.1$ & $0.143437+0.009706i$ & $3.010856-0.402037i$ & $18.376723-0.393336i$\tabularnewline
\cline{2-7} \cline{3-7} \cline{4-7} \cline{5-7} \cline{6-7} \cline{7-7} 
 & \multirow{4}{*}{$10000$} & \multirow{2}{*}{$0.991$} & $0$ & $0$ & $\pm1.930716-0.481345i$ & $\pm13.798347-0.462716i$\tabularnewline
\cline{4-7} \cline{5-7} \cline{6-7} \cline{7-7} 
 &  &  & $0.1$ & $0.127461+0.001895i$ & $2.265562-0.474726i$ & $14.119498-0.462581i$\tabularnewline
\cline{3-7} \cline{4-7} \cline{5-7} \cline{6-7} \cline{7-7} 
 &  & \multirow{2}{*}{$0.996$} & $0$ & $0$ & $\pm3.242616-0.795833i$ & $\pm23.189760-0.764924i$\tabularnewline
\cline{4-7} \cline{5-7} \cline{6-7} \cline{7-7} 
 &  &  & $0.1$ & $0.213619+0.003591i$ & $3.808829-0.781460i$ & $23.733891-0.764259i$\tabularnewline
\hline 
\end{tabular}
\par\end{centering}
\caption{The lowest-lying QNMs $\omega/\kappa_{-}$ for a massless scalar field
of charge $q$ in a near-extremal BI-dS black hole. In the large $b$
limit, i.e., $b=10000$, the values of $\omega/\kappa_{-}$ with $\varLambda=0.02$,
$Q/Q_{\textrm{ext}}=0.991$, $q=0.1$ and $\varLambda=0.06$, $Q/Q_{\textrm{ext}}=0.996$,
$q=0$ are consistent with those in a RN-dS black hole obtained in
\cite{Cardoso:2017soq,Mo:2018nnu}, respectively.}

\label{table-QNM}
\end{table}

In this section, we present the numerical results about the low-lying
QNMs for a scalar field and check the validity of SCC. The results
shown in this sections are obtained with the Mathematica package of
\cite{Jansen:2017oag,Jansen,url-centra.tecnico.} and checked with
some QNMs given in \cite{Breton:2017hwe}, where the WKB approximation
was used. Since SCC may be violated near extremality in a RN-dS black
hole, we here focus on the near-extremal parameter space of a BI-dS
black hole.

In Table \ref{table-QNM}, we show the lowest-lying QNMs $\omega/\kappa_{-}$
of some representative points in the relevant parameter region. Note
that some results in \cite{Cardoso:2017soq,Mo:2018nnu} are recovered
in the large $b$ limit as expected. Besides, when the scalar field
is charged, the symmetry between left and right modes is broken due
to the presence of scalar charge, which was also observed in \cite{Mo:2018nnu,Dias:2018ufh}.
Similar to the RN-dS black hole case, we find that the violation of
SCC occurs when the black hole lies close enough to extremality (e.g.,
$Q/Q_{\textrm{ext}}=0.996$). Interestingly, it shows that a smaller
value of $b$ tends to decrease the absolute value of $\textrm{Im(}\omega)/\kappa_{-}$,
which can alleviate the violation of SCC and even save SCC. Note that
we set $M=16\pi$ without loss of generality in this section.

\subsection{Neutral Scalar Field}

\label{subsec:Massless-Neutral-Scalar}

\begin{figure}
\begin{centering}
\includegraphics{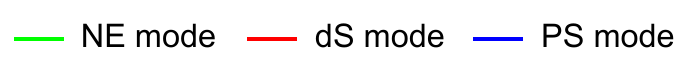}
\par\end{centering}
\begin{centering}
\subfloat[{\small{}Dominant modes of three families for various values of $b$
and $\Lambda$ with varying $Q/Q_{\textrm{ext}}$. The solid vertical
lines indicate the minimal charge ratio$,$ below which a BI-dS black
hole can not possess the Cauchy horizon. Note that the minimal charge
ratio lines in the left column lie out of the relevant region. It
shows that the SCC violation region, in which $\max\left\{ \textrm{Im}(\omega)/\kappa_{-}\right\} <-1/2$,
decreases in size as $b$ decreases.}]{\begin{centering}
\includegraphics[scale=0.43]{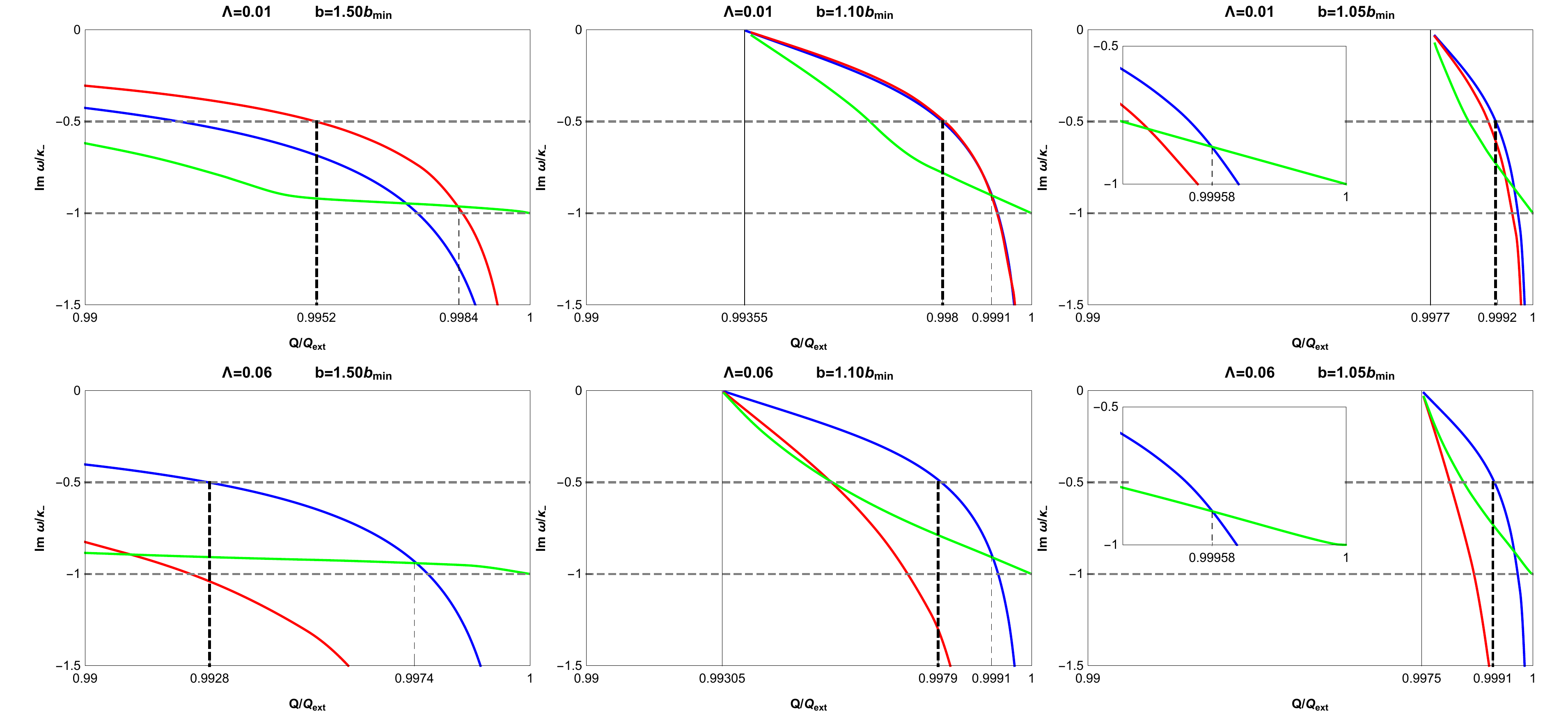}
\par\end{centering}

\label{figure-three-fam-fix-b}}
\par\end{centering}
\begin{centering}
\subfloat[{\small{}Dominant modes of three families for various values of $Q/Q_{\textrm{ext}}$
and $\Lambda$ with varying $b$. The solid vertical lines indicate
the minimal $b$, below which a BI-dS black hole can not possess the
Cauchy horizon. It shows that SCC can be saved for a small enough
value of $b$.}]{\begin{centering}
\includegraphics[scale=0.43]{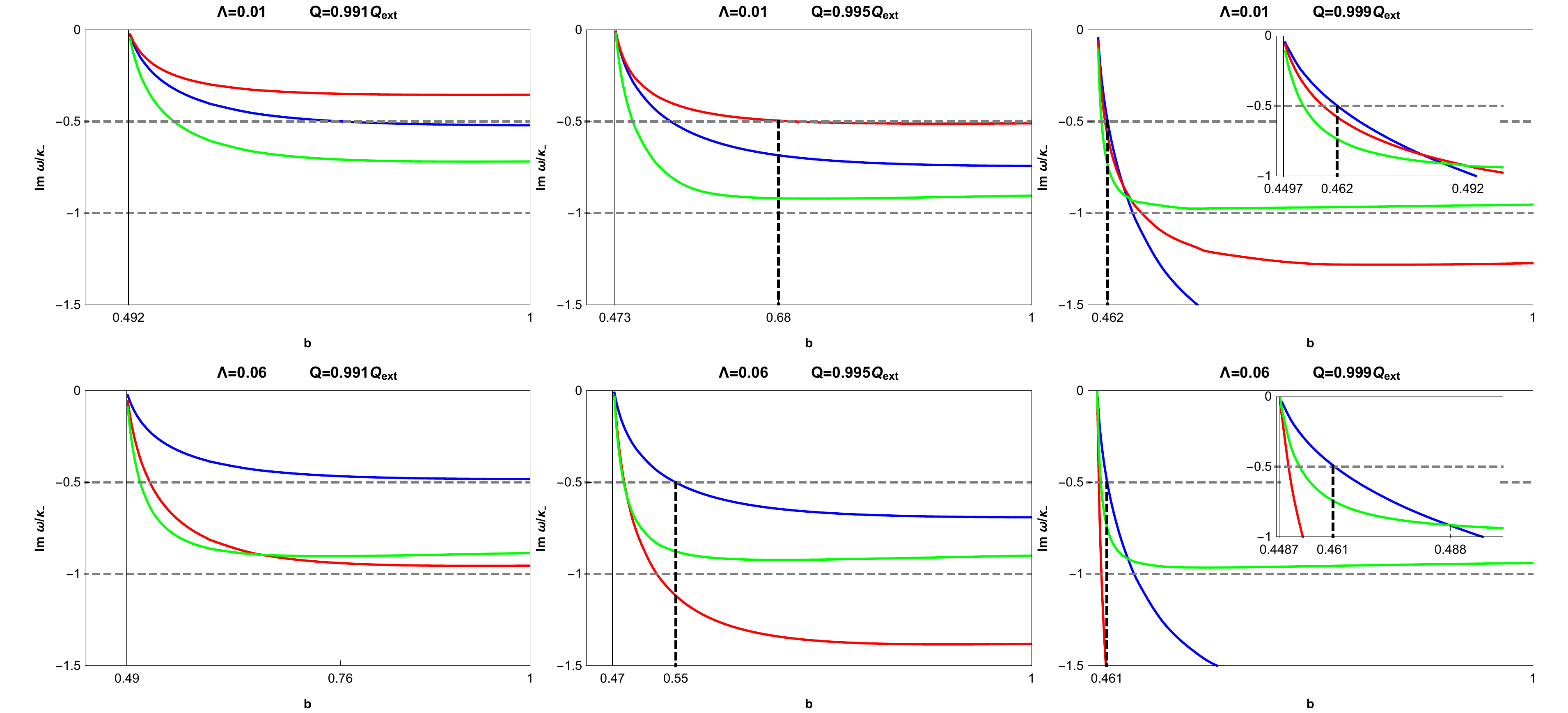}
\par\end{centering}

\label{figure-three-fam-fix-q}}
\par\end{centering}
\caption{Dominant modes of three families for a neutral massless scalar field,
showing the dominant NE mode (green) at $l=0$, the dominant dS mode
(red) at $l=1$ and the (nearly) dominant complex PS mode (blue) at
$l=10$. The thick and thin dashed black vertical lines designate
the points where $\beta\equiv-\textrm{Im}(\omega)/\kappa_{-}=1/2$
and where the NE modes become dominant, respectively.}

\label{figure-three-family}
\end{figure}

\begin{figure}
\begin{centering}
\includegraphics[scale=0.8]{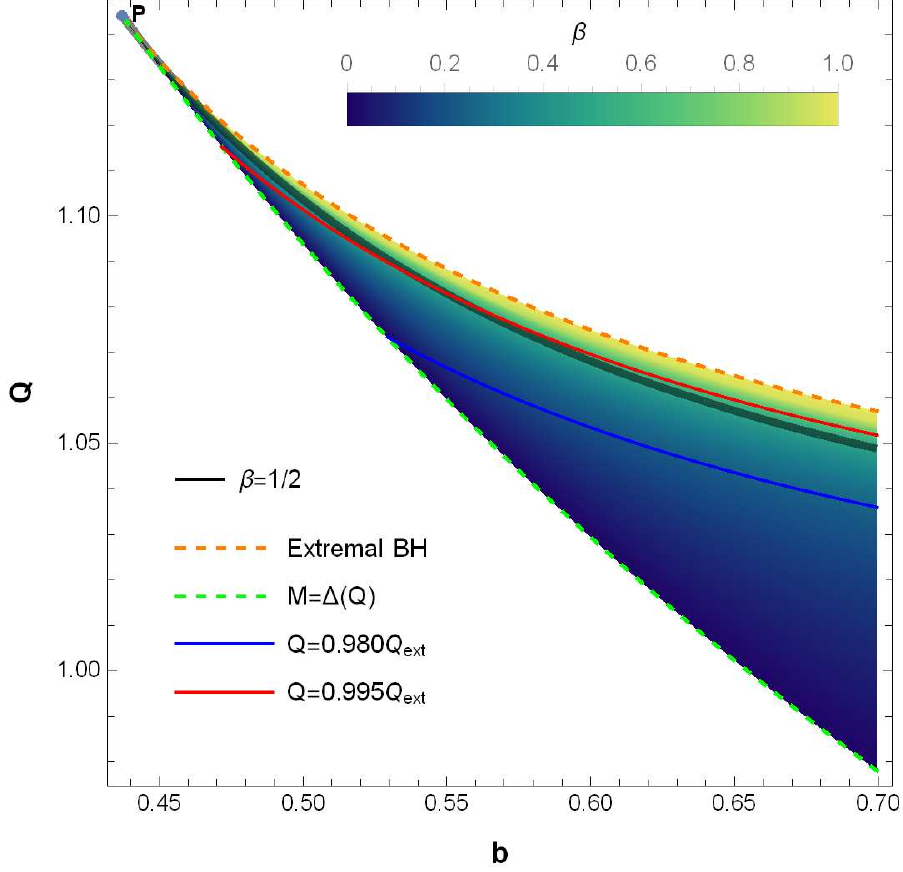}
\par\end{centering}
\caption{Density plot of $\beta$ for a neutral massless scalar field in a
BI-dS black hole with $\Lambda=0.06$. The parameter space of interest
is bounded by $M=\varDelta(Q)$ (dashed green) and the extremal black
hole (dashed orange). Note that $\beta=1/2$ on the thick black line.
So SCC is violated in the region above the thick black line. Two lines
in red and blue denote the near-extremal black holes with the charge
$Q/Q_{\textrm{ext}}=0.995$ and $Q/Q_{\textrm{ext}}=0.980$, respectively.
In the $Q/Q_{\textrm{ext}}=0.995$ case, SCC is violated in some parameter
region but can be saved when $b\text{ is small}$ enough.}

\label{figure-denisty-plot}
\end{figure}

Recently, the authors of \cite{Cardoso:2017soq} found three qualitatively
different families of QNMs for a RN-dS black hole: \textsl{the photon
sphere (PS) family}, which can be traced back to the photon sphere,
\textsl{the de Sitter (dS) family}, which is deformation of the pure
de Sitter modes, and\textsl{ the near-extremal (NE) family}, which
only appears for near-extremal black holes. Similarly, we also observe
these three distinct families for a neutral massless scalar field
in a near-extremal BI-dS black hole. In Fig. \ref{figure-three-family},
we plot the dominant modes of each of the families divided by $\kappa_{-}$.
Specifically, $\textrm{Im(}\omega)/\kappa_{-}\text{ is plotted against }$$Q/Q_{\textrm{ext}}$
for various values of $b$ and $\Lambda$ in Fig. \ref{figure-three-fam-fix-b}.
As shown in Fig. \ref{Region}, for a fixed value of $b$ not far
from $b_{\textrm{min}}$, the $M=\varDelta(Q)$ line puts a lower
bound on $Q/Q_{\textrm{ext}}$, which is depicted as the solid vertical
lines in Fig. \ref{figure-three-fam-fix-b}. It is noteworthy that
all $\textrm{Im}(\omega)/\kappa_{-}$ go to zero as $Q/Q_{\textrm{ext}}$
approaches its lower bound. Indeed, in the limit of $M\rightarrow\varDelta(Q)$,
the Cauchy horizon radius $r_{-}$ goes to zero, and hence the surface
gravity at the Cauchy horizon $\kappa_{-}$ becomes
\begin{equation}
\kappa_{-}=\left|\frac{1}{2r}\frac{d\left(rf\left(r\right)\right)}{dr}\mid_{r=r_{-}}\right|\cong\frac{5bQ}{3r_{-}}\rightarrow\infty,
\end{equation}
where we use eqn. (\ref{eq:limit-fr}). Since QNMs are still finite
in this limit, we find that $\textrm{Im}(\omega)/\kappa_{-}=0$ when
$M=\varDelta(Q)$ (i.e., solid vertical lines in Fig. \ref{figure-three-family}
and dashed green lines in Figs. \ref{Region} and \ref{figure-denisty-plot}).
Moreover, Fig. \ref{figure-three-fam-fix-b} shows that, when $Q/Q_{\textrm{ext}}$
increases towards the extremal limit, $\textrm{the Im}(\omega)/\kappa_{-}$
for the three families' dominant modes all decreases. In the extremal
limit, the PS and dS families become divergent while the NE family
approaches $-1$ from below and hence takes over to make $1/2<\beta<1$.
Thus with fixed values of $b$ and $\Lambda,$ the presence of NE
mode can invalidate SCC as long as the black hole lies close enough
to extremality. Moreover, the dS family is more sensitive to $\Lambda$
than the PS and NE families and can become dominant for ``small''
black holes (small $\Lambda$). Moreover, it shows that the range
of $Q/Q_{\textrm{ext}}$, where SCC is violated, shrinks with decreasing
value of $b$ ($b$ decreases from the left column to the right column
in Fig. \ref{figure-three-fam-fix-b}). To better illustrate the dependence
of $\textrm{Im}(\omega)/\kappa_{-}$ on $b$, we plot $\textrm{Im}(\omega)/\kappa_{-}\text{ against \ensuremath{b} }$for
various values of $Q/Q_{\textrm{ext}}$ and $\Lambda$ in Fig. \ref{figure-three-fam-fix-q}.
It is expected that SCC is easier to be violated when the black hole
is closer to extremality. In fact, increasing $Q/Q_{\textrm{ext}}$
towards extremality from the left column to the right column in Fig.
\ref{figure-three-fam-fix-q}, we find that the SCC violation ranges
of $b$, which are on the left of the dashed vertical lines, increase
in size. Note that there is no SCC violation in the $Q/Q_{\textrm{ext}}=0.991$
case. In Fig. \ref{Region}, it shows that the $Q/Q_{\textrm{ext}}$-constant
line (e.g., the blue line with $Q/Q_{\textrm{ext}}=0.900$) always
has a lower bound $b_{Q/Q_{\textrm{ext}}}$ on $b\text{, which is also imposed by the }$$M=\varDelta(Q)$
line. The solid vertical lines in Fig. \ref{figure-three-fam-fix-q}
represent $b=b_{Q/Q_{\textrm{ext}}},$ on which $M=\varDelta(Q)$
and thus $\beta=0$. So with fixed values of $Q/Q_{\textrm{ext}}$
and $\Lambda$, one can always have $\beta<1/2$ when $b$ is close
enough to $b_{Q/Q_{\textrm{ext}}}$. In the $Q/Q_{\textrm{ext}}=0.995$
and $Q/Q_{\textrm{ext}}=0.999$ cases, SCC is violated for large enough
values of $b$. Nevertheless, we can recover SCC by making $b$ close
enough to $b_{Q/Q_{\textrm{ext}}}$.

Finally, we depict the density plot of $\beta$ in the small $b$
region in Fig. \ref{figure-denisty-plot}, where the solid black line
represents the threshold $\beta=1/2$. So SCC is violated in the region
between the extremal line (dashed orange) and $\beta=1/2$ (solid
black). The $Q/Q_{\textrm{ext}}=0.995$ line is displayed as a red
line, which also shows that SCC can be recovered for a small enough
value of $b$$.$ For a near-extremal BI-dS black hole with a constant
charge $Q$, it also shows in Fig. \ref{figure-denisty-plot} that
SCC is respected when $b$ is close enough to the dashed green line.
Furthermore, Fig. \ref{figure-denisty-plot} displays that SCC can
be recovered for a highly extremal BI-dS black hole as long as the
Born-Infeld parameter $b$ is sufficiently close to $b_{\textrm{min}}$
(the point $P$).

\subsection{Charged Scalar Field}

\label{subsec:Massive-Charged-Scalar}

\begin{figure}
\begin{centering}
\includegraphics[scale=0.6]{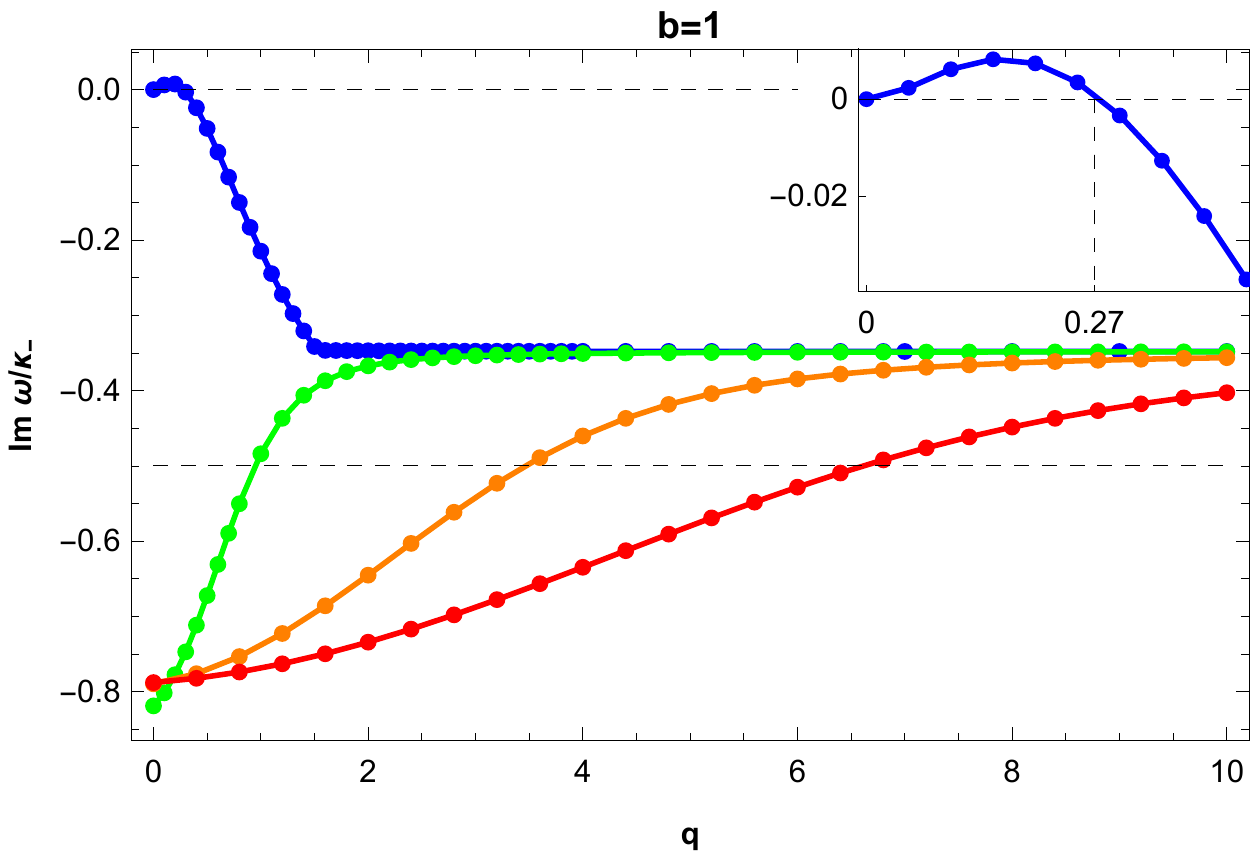}\includegraphics[scale=0.6]{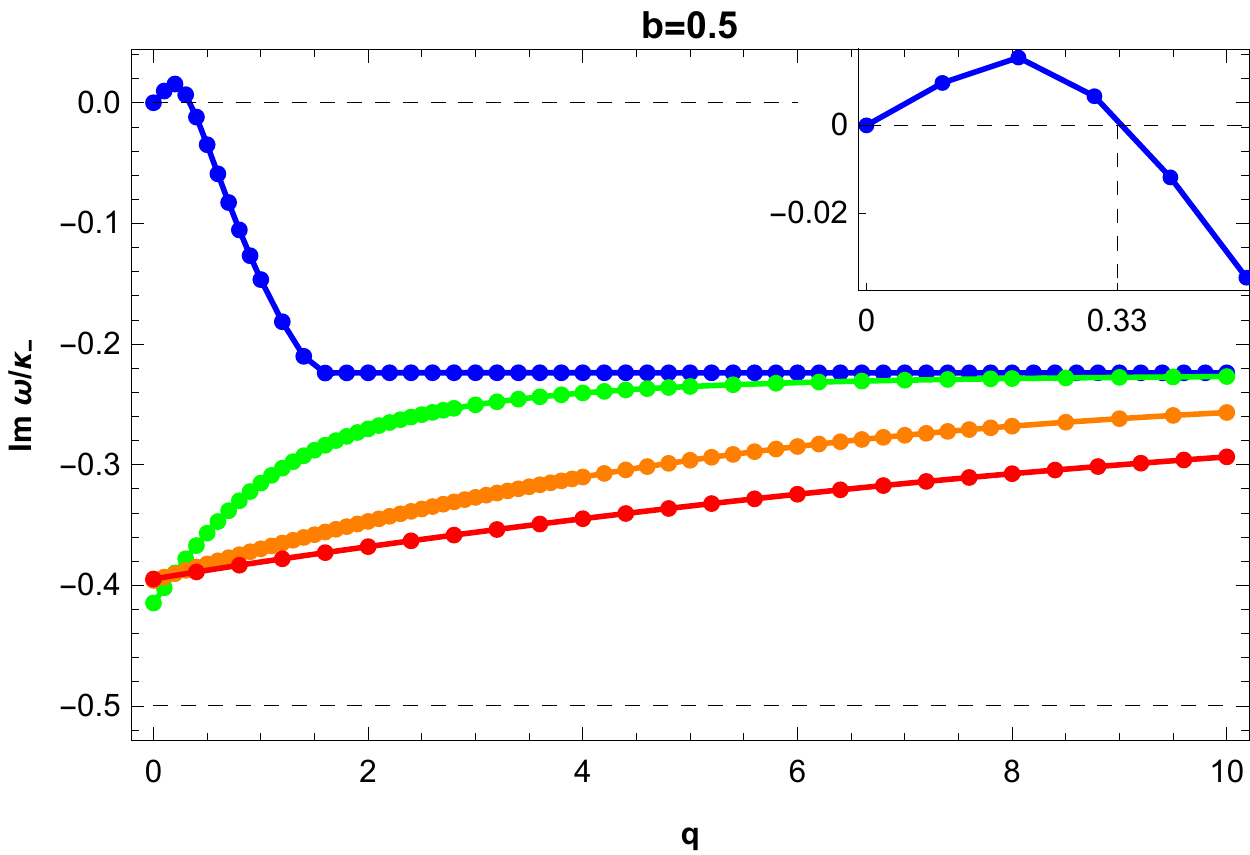}
\par\end{centering}
\begin{centering}
\includegraphics{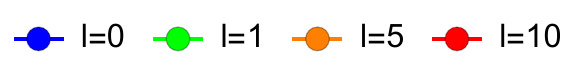}
\par\end{centering}
\caption{Plots of the lowest-lying QNMs as a function of the scalar charge
$q$ for a massless charged scalar field in a BI-dS black hole with
$\varLambda=0.06$ and $Q/Q_{\textrm{ext}}=0.996$. \textbf{Left:}
$b=1$. \textbf{Right:} $b=0.5$. The dashed vertical line designates
an upper bound on the scalar charge, under which superradiance is
present.}

\label{figure-q-vary}
\end{figure}

We now turn on the charge of a scalar field and investigate the validity
of SCC. In Fig. \ref{figure-q-vary}, we first plot the lowest-lying
QNMs as a function of the scalar charge $q$ for a massless charged
scalar field in a BI-dS black hole, which behave rather similarly
to the RN-dS black hole case. The blue lines represent the $l=0$
zero mode, which reduces to a trivial mode in the limit $q\rightarrow0$.
In particular, we observe the presence of superradiant instability
in the small scalar charge regime. This linear instability suggests
that the perturbations will be severely unstable even in the exterior
of the black hole, and thus one can not infer anything about SCC when
superradiance occurs. Note that the non-smoothness of the blue lines
around $q\sim1.5$ is caused by the competition between the PS and
NE modes. The $b=1$ case is shown in the left panel of Fig. \ref{figure-q-vary},
which shows that SCC is violated for $q=0$ since the $l=0$ zero
trivial mode is discarded. However for a nonzero $q$, SCC is saved
out due to the nontrivial $l=0$ zero mode. The right panel of Fig.
\ref{figure-q-vary} displays that, for a smaller value of $b=0.5$,
the higher $l$-modes are also above the threshold line $\beta=1/2$,
and the superradiant regime increases in size, which means that small
$b$ tends to make the black hole more unstable. From Fig. \ref{figure-q-vary},
we see that $\beta$ is determined by the $l=0$ dominant mode for
a charged scalar field.

\begin{figure}
\begin{centering}
\includegraphics[scale=0.9]{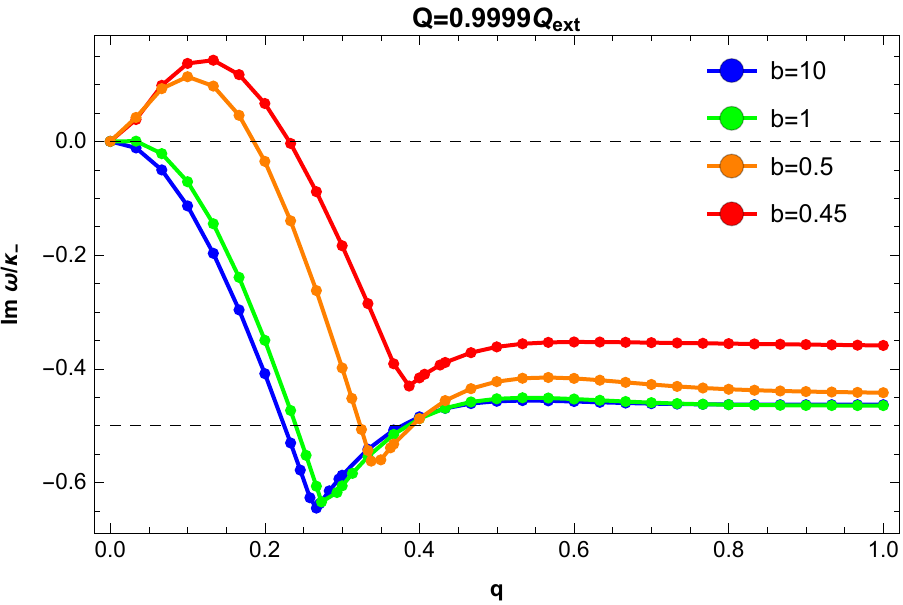}\includegraphics[scale=0.9]{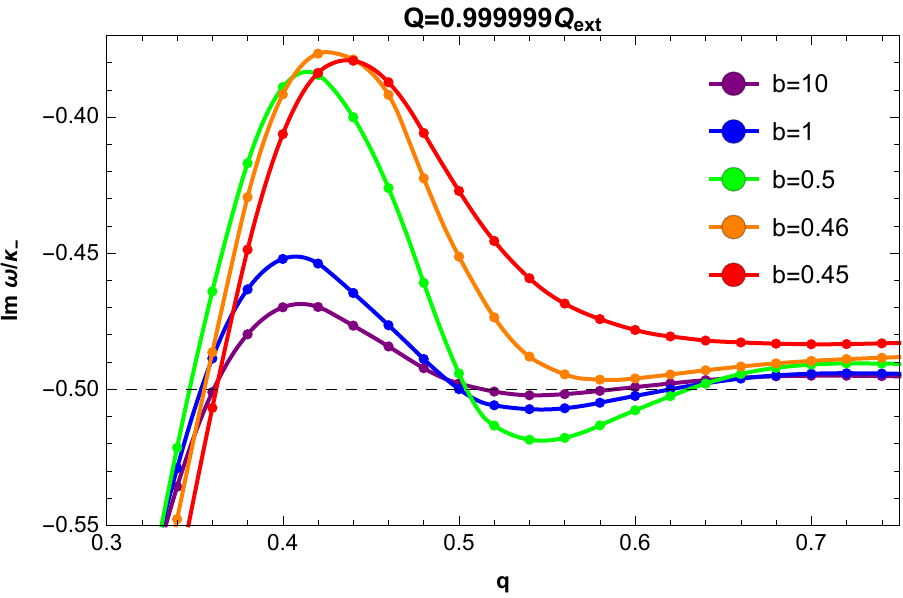}
\par\end{centering}
\caption{Plots of the $l=0$ dominant mode as a function of the scalar charge
$q$ for a massless charged scalar field in a BI-dS black hole with
$\Lambda=0.14$.\textbf{ Left:} $Q/Q_{\textrm{ext}}=1-10^{-4}$. The
SCC violation region decreases in size and even disappears with decreasing
$b$. \textbf{Right: }$Q/Q_{\textrm{ext}}=1-10^{-6}$. The presence
of the wiggles for large values of $b$ ($b=10,1\textrm{ and }0.5$)
and absence of the wiggles for small values of $b$ ($b=0.46\textrm{ and }0.45$)
are observed .}

\label{figure-violation-and-wiggle}
\end{figure}

Further increasing the black hole charge $Q$ towards extremality,
we plot the $l=0$ dominant mode as a function of the scalar charge
$q$ for a massless charged scalar field in a BI-dS black hole in
Fig. \ref{figure-violation-and-wiggle}. The dependence of the $l=0$
dominant mode on $b$ is plotted in the left panel of Fig. \ref{figure-violation-and-wiggle},
where $Q/Q_{\textrm{ext}}=1-10^{-4}$. The curve with $b=10$ is almost
identical to the RN-dS case, which was shown in Fig.11 of \cite{Mo:2018nnu}.
It shows that the SCC violation occurs for $b=10,$ $1$ and $0.5$
in some scalar charge regime. Nevertheless, these violation regions
decrease in size as $b$ decreases. Interestingly, SCC is always respected
when $b=0.45.$ In the right panel of Fig. \ref{figure-violation-and-wiggle},
we plot the $l=0$ dominant mode for a more extremal BI-dS black hole
with $Q/Q_{\textrm{ext}}=1-10^{-6}$ and display that the ``wiggles'',
i.e., small oscillations around $\beta=1/2$, appear in the $b=10,1\textrm{ and }0.5$
cases. It is noteworthy that the wiggles disappear, and SCC is restored
when $b\lesssim0.46$.

\begin{figure}
\begin{centering}
\includegraphics[scale=0.45]{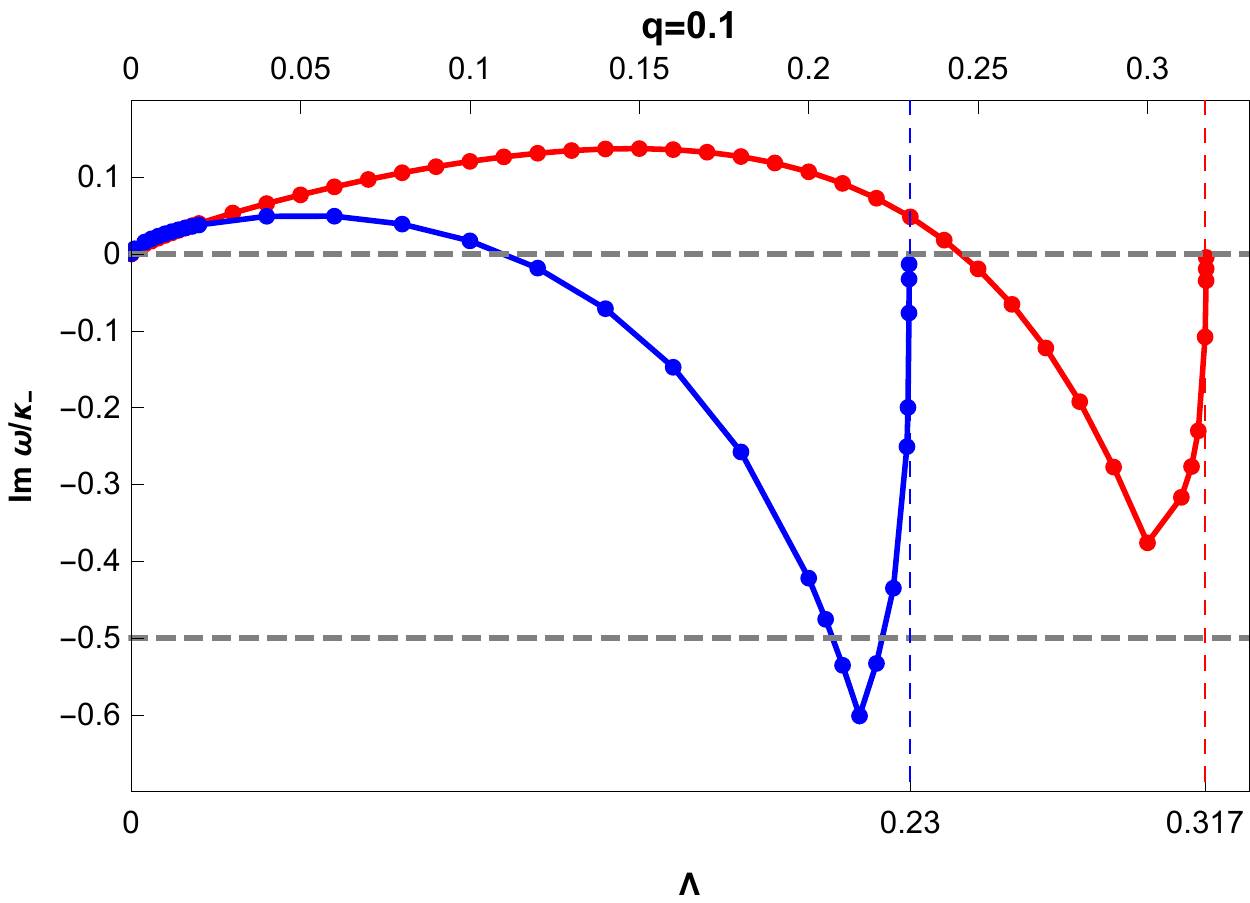}\includegraphics[scale=0.45]{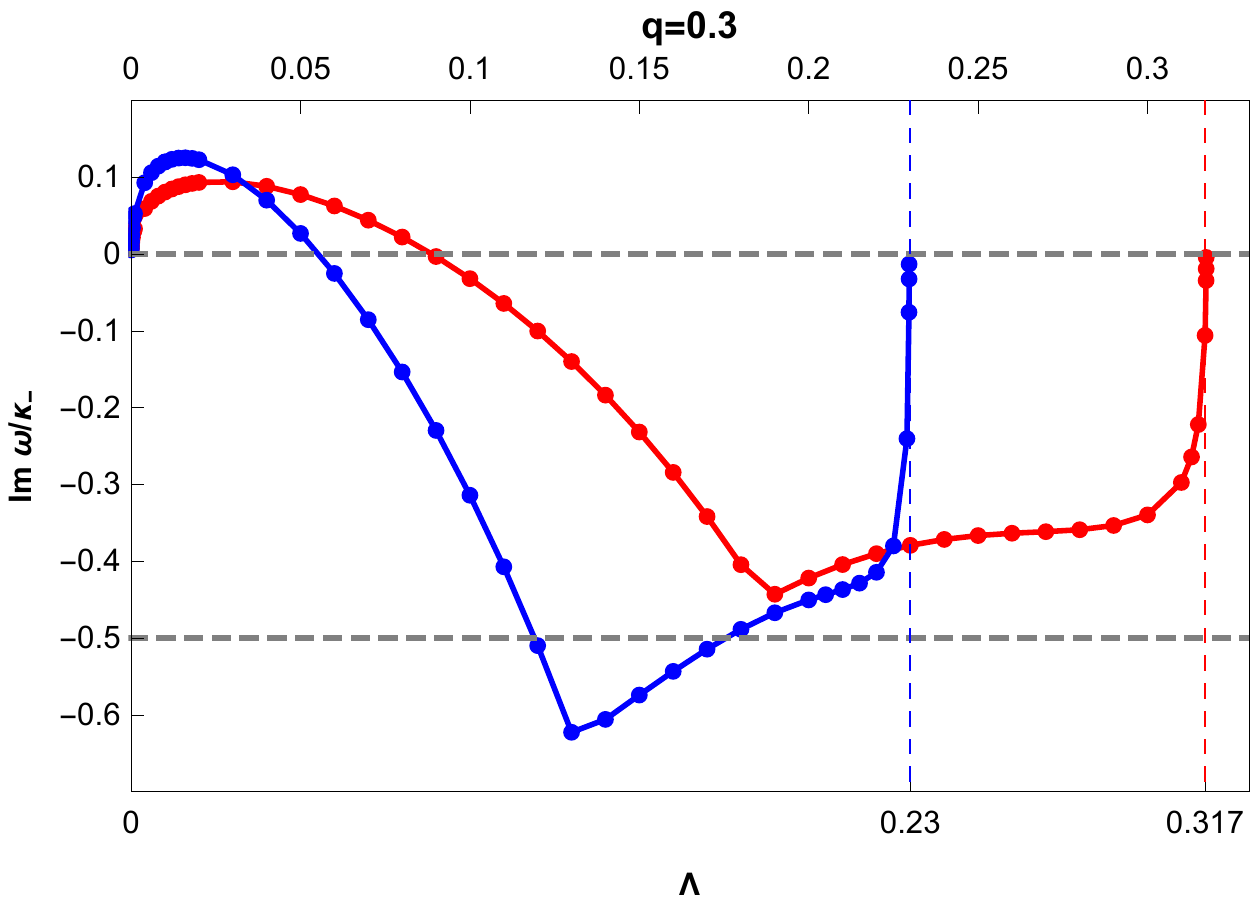}\includegraphics[scale=0.45]{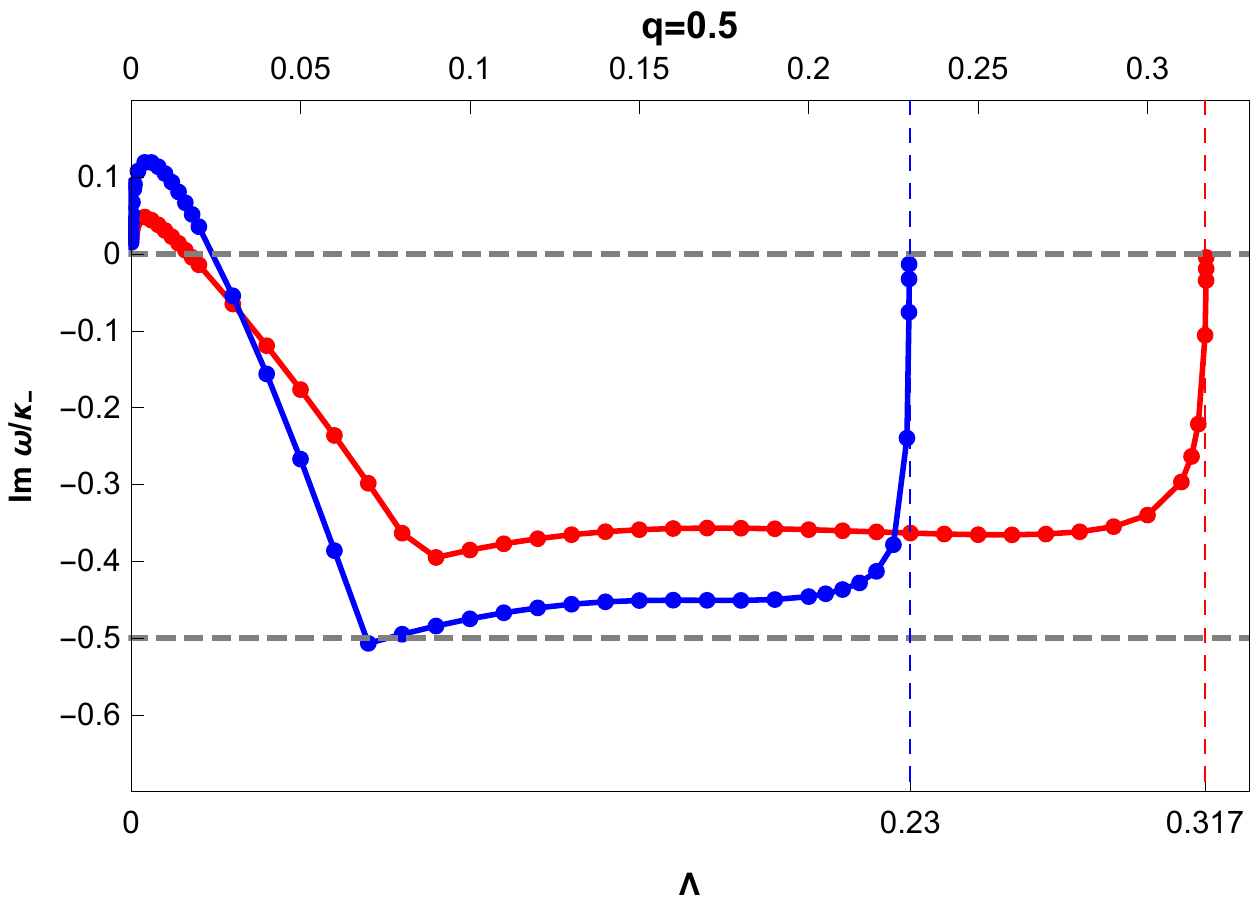}
\par\end{centering}
\begin{centering}
\includegraphics{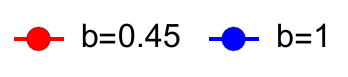}
\par\end{centering}
\caption{Plots of the $l=0$ dominant mode for a massless charged scalar field
as a function of the cosmological constant $\Lambda$ in a BI-dS black
hole with $Q/Q_{\textrm{ext}}=1-10^{-4}$ for the scalar charge $q=0.1$,
$0.3$ and $0.5$. The dashed vertical blue and red lines represent
the maximum values of $\Lambda$ for $b=1$ and $b=0.45$, respectively.
SCC is respected for the smaller $b$ (i.e., $b=0.45$) while there
exists the SCC violation region for the larger $b$ (i.e., $b=1$).}

\label{figure-lambda-vary}
\end{figure}

Next we turn to the dependence of the $l=0$ dominant mode on $\varLambda$
in Fig. \ref{figure-lambda-vary}, where the $l=0$ dominant mode
is plotted as a function of $\Lambda$ for various values of $q$
and $b$. Note that $\Lambda$ is bounded above by a maximum value
due to the Nariai limit of a BI-dS black hole. In the Nariai limit,
$\textrm{Im(}\omega)$ approaches zero while $\kappa_{-}$ stays finite,
which explains $\textrm{Im}(\omega)/\kappa_{-}=0$ at the maximum
value of $\Lambda$ shown in Fig. \ref{figure-lambda-vary}. Again,
the non-smoothness in Fig. \ref{figure-lambda-vary} results from
the competition between the PS and NE modes. When $b=1$, SCC can
be violated in some parameter region of $\Lambda$ and $q$. For a
smaller $b=0.45$, these violation regions all disappear.

\begin{figure}
\begin{centering}
\includegraphics[scale=0.9]{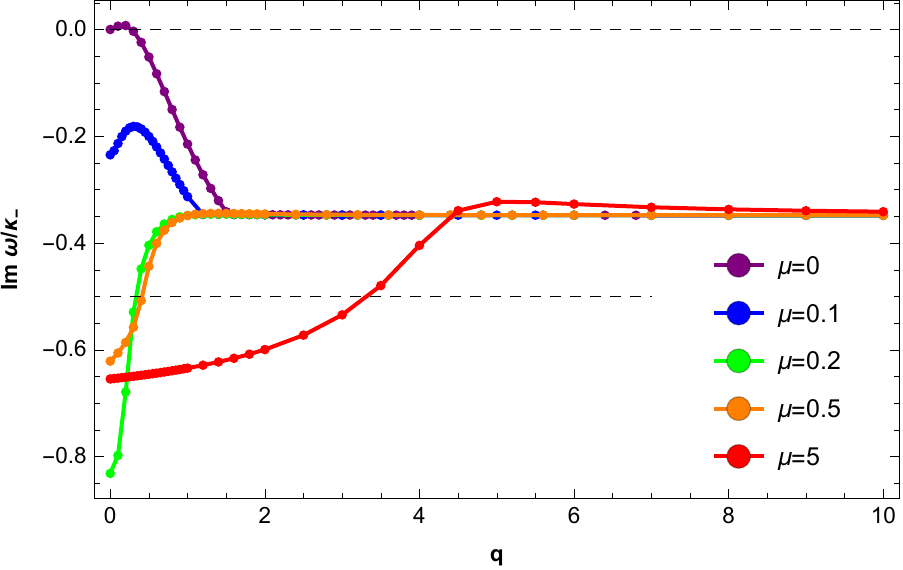}\includegraphics[scale=0.9]{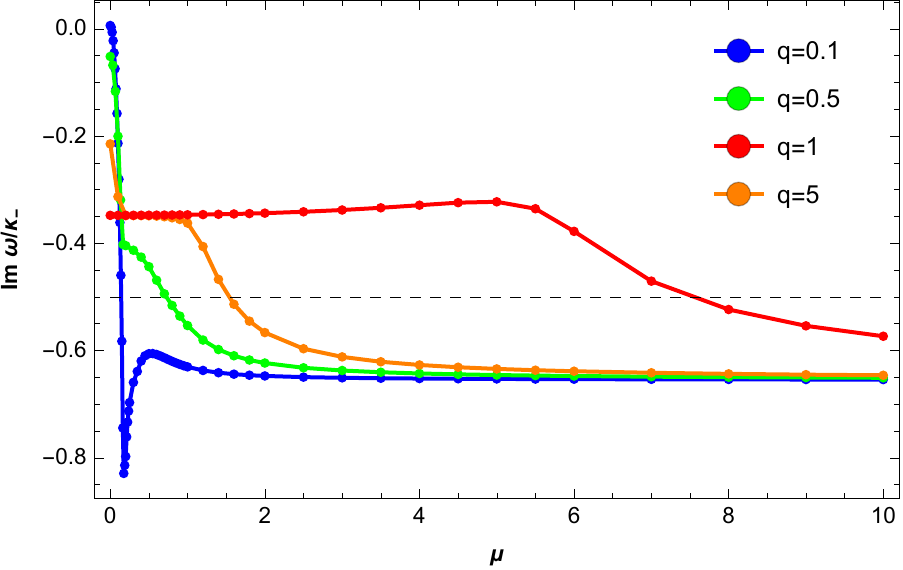}
\par\end{centering}
\caption{Plots of the $l=0$ dominant mode for a massive charged scalar field
of charge $q$ and mass $\mu$ in a BI-dS black hole with $\varLambda=0.06$,
$Q/Q_{\textrm{ext}}=0.996$ and $b=1$. \textbf{Left:} As a function
of $q$ for various values of $\mu$. \textbf{Right:} As a function
of $\mu$ for various values of $q$. It shows that SCC is violated
more easily for a more massive scalar.}

\label{figure-massive}
\end{figure}

Finally, we investigate the dependence of the $l=0$ dominant mode
on the scalar mass $\mu$ in Fig. \ref{figure-massive}. As shown
in the left panel in Fig. \ref{figure-massive}, the superradiant
instability is highly sensitive to the scalar mass. For a sufficiently
large value of $\mu$, superradiant instability no longer exists.
Moreover, the $\textrm{Im}(\omega)/\kappa_{-}$ of the dominant $l=0$
mode can be smaller than $-1/2$ for large enough $\mu$. It also
displays, in the right panel of Fig.\ref{figure-massive}, that the
$l=0$ dominant mode for various values of $q$ all go below the threshold
line when the scalar field is sufficiently massive. So Fig. \ref{figure-massive}
shows that SCC tends to be violated for a larger scalar mass. On the
other hand, it also shows that SCC tends to be saved for a larger
scalar charge.

\section{Conclusion}

\label{sec:Discussion-and-conclusion}

In this paper, we investigated the validity of SCC in a BI-dS black
hole perturbed by a scalar field with/without a charge. After the
parameter region, where a BI-dS black hole can possess the Cauchy
horizon, was obtained in Section \ref{sec:BI-dS-black-hole}, we presented
the numerical results for a neutral scalar field in Section \ref{subsec:Massless-Neutral-Scalar}
and a charged scalar in Section \ref{subsec:Massive-Charged-Scalar}.

For the Born-Infeld parameter $b\gtrsim1$, the behavior of SCC in
a BI-dS black hole is quite similar to that in a RN-dS black hole.
In fact, we observed that SCC is always violated when a BI-dS black
hole is sufficiently close to extremality in the neutral case, and
the SCC violation region, especially the wiggles, can appear in the
charged case. On the other hand, for a smaller value of $b$, the
NLED effect can play an important role and tends to alleviate the
violation of SCC. Specifically, we found that
\begin{itemize}
\item For a massless neutral scalar field, Fig. \ref{figure-three-family}
showed that the SCC violation region decreases in size with deceasing
$b$.
\item For a massless neutral scalar field, Fig. \ref{figure-denisty-plot}
showed that SCC can always be restored for a near-extremal BI-dS black
hole with a fixed charge ratio $Q/Q_{\text{ext}}$ or a charge $Q$
when $b$ is sufficiently small.
\item For a massless charged scalar, Figs. \ref{figure-violation-and-wiggle}
and \ref{figure-lambda-vary} showed that the SCC violation region
also decreases in size with deceasing $b$. Moreover, the violation
region can disappear for a sufficiently small value of $b$.
\end{itemize}
The dependence of SCC on the scalar mass and charge was discussed
in Fig. \ref{figure-massive} for a massive charged scalar, which
showed that the smaller the scalar mass is (or the larger the scalar
charge is), the easier it becomes to restore SCC. Our results indicate
that the quantum effects could play a crucial role in rescuing SCC.
Therefore, it is inspiring to check the validity of SCC in modified
gravity theories, even a quantum gravity model.

\vspace{5mm}

\noindent {\bf Acknowledgements} 
We are deeply grateful to Hongbao Zhang for his excellent presentations and valuable comments. We also  thank Haitang Yang for his helpful discussions and suggestions. This work is supported in part by NSFC (Grant No. 11005016, 11875196 and 11375121).

\end{document}